# Atomically thin group-V elemental films: theoretical investigations of antimonene allotropes

Gaoxue Wang[1], Ravindra Pandey[1]*, and Shashi P. Karna[2]

[1]Department of Physics, Michigan Technological University, Houghton, Michigan 49931, USA

[2]US Army Research Laboratory, Weapons and Materials Research Directorate, ATTN: RDRL-WM, Aberdeen Proving Ground, MD 21005-5069, USA

(May 7, 2015)

*Email: pandey@mtu.edu
shashi.p.karna.civ@mail.mil

# ABSTRACT

Group-V elemental monolayers including phosphorene are emerging as promising 2D materials with semiconducting electronic properties. Here, we present the results of first principles calculations on stability, mechanical and electronic properties of 2D antimony (Sb), antimonene. Our calculations show that free-standing $α$ and $β$ allotropes of antimonene are stable and semiconducting. The $α$-Sb has a puckered structure with two atomic sub-layers and $β$-Sb has a buckled hexagonal lattice. The calculated Raman spectra and STM images have distinct features thus facilitating characterization of both allotropes. The $β$-Sb has nearly isotropic mechanical properties while $α$-Sb shows strongly anisotropic characteristics. An indirect-direct band gap transition is expected with moderate tensile strains applied to the monolayers, which opens up the possibility of their applications in optoelectronics.

**INTRODUCTION**

Group-V elemental monolayers have recently emerged as novel two dimensional (2D) materials with semiconducting electronic properties. For example, the monolayer form of black phosphorous, phosphorene (α-P), has a direct band gap and high carrier mobility [1-2], which can be exploited in the electronics [3-4]. Additionally, the stability of phosphorene in the other allotropes including β, γ, and δ phases was predicted [5-6]. The equilibrium configuration of α-P is puckered due to the intralayer $sp^3$ bonding character in the lattice. The 2D form of the so-called blue phosphorus is referred to as β-P [5] which possesses the hexagonal honeycomb structure maintaining $sp^3$ character of bonds. Each atom is three-fold coordinated forming silicene-like 2D structure with buckling at the surface [7]. γ-P and δ-P have rectangular Wigner-Seitz cells [6].

Considering the chemical similarity of elements belonging to the same column in the periodic table, the other group-V elemental monolayers have also been investigated (Table S1, Supporting Information). Arsenene in α and β phases is predicted to be stable [8-10]. Ultrathin Bi (111) and Bi (110) films have been assembled on Si substrate or pyrolytic graphite in experiments [11-16]. It is important to note that, unlike group-IV monolayers which are semi-metallic including graphene [17], silicene [7], and germanene [18], group-V monolayers are found to be semiconductors [8-10, 15], thereby offering prospects for device applications at nanoscale.

In bulk, several allotropes exist for group-V elements at ambient conditions. For example, the most stable allotrope for P is black phosphorus which is composed of AB stacked α-P monolayers. It possesses an intrinsic band gap of ~0.3 eV [1, 19] which increases to ~2 eV in its monolayer form [20]. The other group-V elements, As, Sb, and Bi, crystallize in a rhombohedral structure at ambient conditions, where the (111) direction is composed of ABC stacked β- phase monolayers [21].

In this paper, we focus on the 2D antimony (Sb), referred to as antimonene. Recently, Zhang et al. have shown that the Sb (111) films (i.e. β-Sb) undergo a thickness dependent transition from topological semimetal to topological insulator to normal semiconductor with decreasing thickness [21]. The semiconducting electronic properties of *β*-Sb monolayer is also confirmed by a recent theoretical investigation [22]. However, stability and electronic properties of antimonene in other allotropes (i.e. *α*, *γ*, and *δ*-Sb) have not yet been investigated.

We consider antimonene allotropes including *α*-, *β*-, *γ*-, and *δ*-Sb examining their stability by phonon dispersion calculations based on density functional theory (DFT). Furthermore, we will investigate the effect of mechanical strain on the electronic properties of antimonene allotropes. We will also calculate Raman spectra and scanning tunneling microscope (STM) images to gain further insights into the electronic structure and surface morphology of antimonene.

**COMPUTATIONAL DETAILS**

The calculations were performed with the use of VASP program package [23]. We employed the local density approximation (LDA) together with the projector-augmented-wave (PAW) [24] method which has been shown to correctly describe Sb films [21]. For bulk Sb, the calculated lattice constant of 4.31 Å is in excellent agreement with the experimental value of 4.30 Å [25] giving confidence in the proposed approach based on the LDA-DFT level of theory. To compare stability and structural parameters of different allotropes of antimonene, the Perdew-Burke-Ernzerhof (PBE) [26] functional and the DFT-D2 method of Grimme [27] were also employed.

In calculations, the energy convergence was set to $10^{-6}$ eV and the residual force on each atom was smaller than 0.01 eV/Å. The cutoff energy for the plane-wave basis was set to 500 eV. The reciprocal space was sampled by a grid of (15×15×1) *k* points in the Brillouin zone. The vacuum distance normal to the plane was larger than 20 Å to eliminate interaction between the replicas due to the period boundary conditions in the supercell approach of our model. The spin-orbit coupling (SOC) was included in calculations for the band structure. The Phonopy code [28] was used for the phonon dispersion calculation considering supercell of (4×5) for *α*-Sb, (5×5) for *β*-Sb, (5×4) for *γ*-Sb, and (3×3) for *δ*-Sb. The non-resonance Raman spectra were obtained within density-functional perturbation theory (DFPT) by second order response to an electric field as implemented in Quantum Espresso [29]. The scanning tunneling microscope (STM) images are based on BTH approximation [30], which has been successfully used to investigate tunneling characteristics of nanostructures including BN monolayer and phosphorene [31-34].

## RESULTS AND DISCUSSIONS

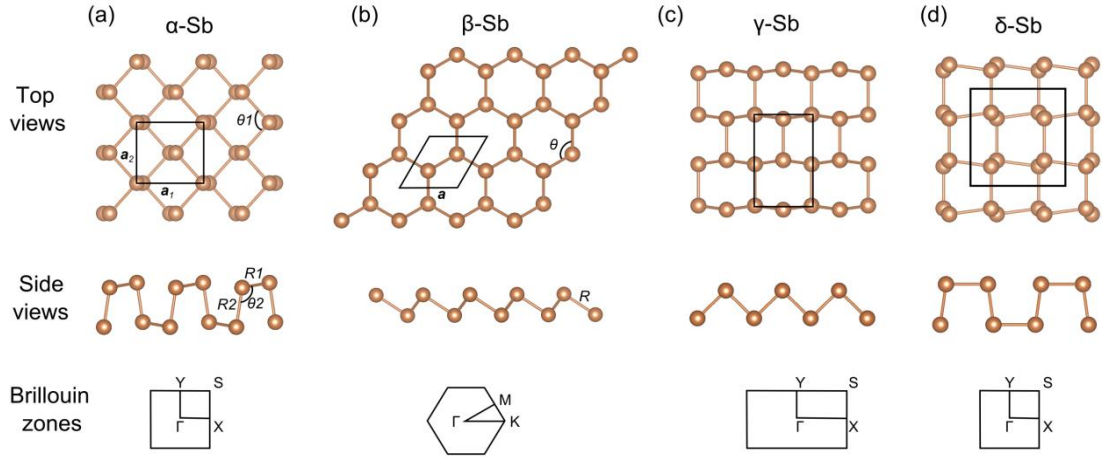

*Figure 1. The structural configurations of antimonene allotropes: (a) α-Sb, (b) β-Sb, (c) γ-Sb, and (d) δ-Sb.*

The structural configurations of antimonene allotropes are shown in Figure 1. The $α$-Sb has a distorted atomic structure with two sub-layers, where atoms belonging to the same sub-layer are not in the same planes (Figure 1(a)). The four atoms in the unit cell are arranged in a rectangular lattice with a puckered surface. The calculated bond lengths are 2.83 and 2.91 Å and the calculated bond angles are 95.0 and 102.5° at LDA-DFT level of theory for $α$-Sb (Table 1).

The ground state configuration of $β$-Sb mimics the metallic Sb (111) surface (Figure 1(b)). It has a hexagonal lattice with the buckled surface similar to what was predicted for $β$-P. The bond length between neighboring Sb atoms is 2.84 Å, and the bond angle is 89.9° (Table 1). The results are in agreement with previous theoretical calculations on $β$-Sb monolayer [22, 35]. Similar to $γ$- and $δ$-P [6], the $γ$- and $δ$-Sb have the rectangular unit cells which are shown in Figures 1(c) and 1(d). The calculated bond lengths are (2.82 and 2.94 Å) and (2.87 and 2.93 Å) for γ-Sb, and δ-Sb, respectively.

The stability of these antimonene allotropes is first investigated by the calculation of the phonon dispersion curves as shown in Figure 2. No imaginary vibrating mode is seen for α-Sb and β-Sb illustrating their stability as the free-standing monolayers. The phonon dispersion curve of β-Sb is similar to that of phosphorene with separated acoustic and optical modes. The maximum vibrational frequency in α-Sb and β-Sb is 170 and 200 cm$^{-1}$, respectively. Our calculations show that γ-Sb has imaginary mode along Γ-X, and δ-Sb has imaginary modes at Γ. Employing a larger supercell model with a higher convergence criteria also yielded imaginary frequencies for γ- and δ-Sb, thus confirming their structural instabilities.

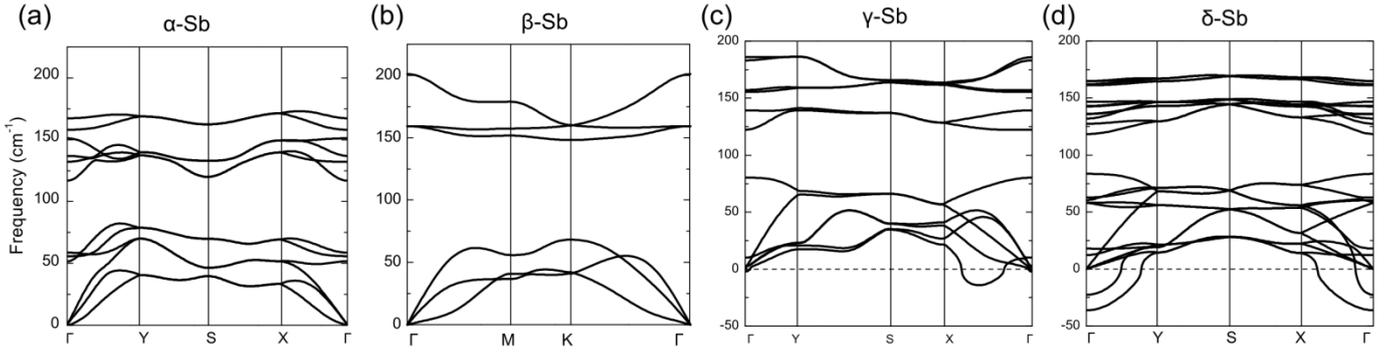

Figure 2. The calculated phonon dispersions of antimonene allotropes: (a) α-Sb, (b) β-Sb, (c) γ-Sb, and (d) δ-Sb.

The stability of α- and β-Sb monolayers is further examined by the cohesive energy calculations using several exchange and correlational functional forms of DFT. As listed in Table 1, α-Sb has larger cohesive energy than β-Sb at the LDA-DFT (≈60 meV) and DFT-D2 (≈30 meV) levels of theory, though both monolayers are nearly degenerate at GGA(PBE)-DFT level of theory.

Table 1. The ground state structural parameters (see Figure 1) of antimonene allotropes: a is *the lattice constant, R is the near-neighbor distance, θ is the bond angle, and Ec is the cohesive energy which is taken to be the total energy difference between the 2D material and its constituting atoms.*

| Level of theory | α-Sb | | | | | | | β-Sb | | | |
|---|---|---|---|---|---|---|---|---|---|---|---|
| | $a_1$ (Å) | $a_2$ (Å) | $R_1$ (Å) | $R_2$ (Å) | $\theta_1$ (°) | $\theta_1$ (°) | $E_c$ (eV/atom) | $a$ (Å) | $R$ (Å) | $\theta$ (°) | $E_c$ (eV/atom) |
| LDA | 4.48 | 4.31 | 2.83 | 2.91 | 95.0 | 102.5 | -4.63 | 4.01 | 2.84 | 89.9 | -4.57 |
| GGA(PBE) | 4.74 | 4.36 | 2.87 | 2.94 | 95.3 | 102.4 | -4.03 | 4.12 | 2.89 | 90.8 | -4.03 |
| DFT-D2 | 4.77 | 4.28 | 2.86 | 2.91 | 94.6 | 103.5 | -4.29 | 4.04 | 2.87 | 89.6 | -4.26 |

The phonon free energy difference in the temperature range of 0-600 K (Figure S1, Supporting Information) is calculated to be less than 15 meV/atom between α-Sb and β-Sb suggesting stabilization of both monolayers in experiments. Interestingly, a crossover in the cohesive energies of α-Sb and β-Sb multilayers is predicted which suggests that β-Sb is more stable than α-Sb in multilayers for thickness more than 3 atomic layers (Figure S2, Supporting Information). The thickness dependent phase transition is mainly due to the stronger interlayer interaction in β multilayers (as will be shown later), resulting into their stability over α-Sb multilayers. It may be noticed that the experimental results on ultrathin Bi films show stability of β-Bi over α-Bi for films with thickness more than 4 atomic layers [15].

Considering that the Raman measurements are generally used to characterize 2D materials, such as graphene [36], we have calculated the Raman spectra for α- and β-Sb monolayers shown in Figure 3. In order to assess the reliability of our approach, we first calculated the Raman spectrum of bulk Sb. Two Raman peaks, $E_g$ at ~100 cm$^{-1}$ and $A_{1g}$ at ~148 cm$^{-1}$, were seen (Figure S3, Supporting Information) which are in agreement with experiments [37]. This gives confidence in our calculated LDA-DFT results for the Raman spectra of antimonene.

α-Sb belongs to $C_{2v}$ group, and the modes, $A_1^1$ at 63 cm$^{-1}$, $B_1$ at 102 cm$^{-1}$, $A_1^2$ at 132 cm$^{-1}$, and $A_1^3$ at 147 cm$^{-1}$, exhibit prominent Raman scattering. $A_1^1$ and $A_1^3$ are out-of-plane modes. For the $A_1^1$ mode, atoms belonging to the same sub-layer vibrate along opposite directions. $A_1^3$ is the most dominating Raman peak for α-Sb for which atoms belonging to the same sub-layer vibrate along the same direction and the two sub-layers vibrate opposite to each other. $B_1$ and $A_1^2$ are both in-plane modes in α-Sb. The β-Sb monolayer belongs to $D_{3d}$ group and the Raman active modes are at 150 cm$^{-1}$ ($E_g$) and 195 cm$^{-1}$ ($A_{1g}$). The $E_g$ modes are doubly degenerate in-plane modes with two atoms in the unit cell vibrating along opposite directions, and $A_{1g}$ is an out-of-plane vibrating mode.

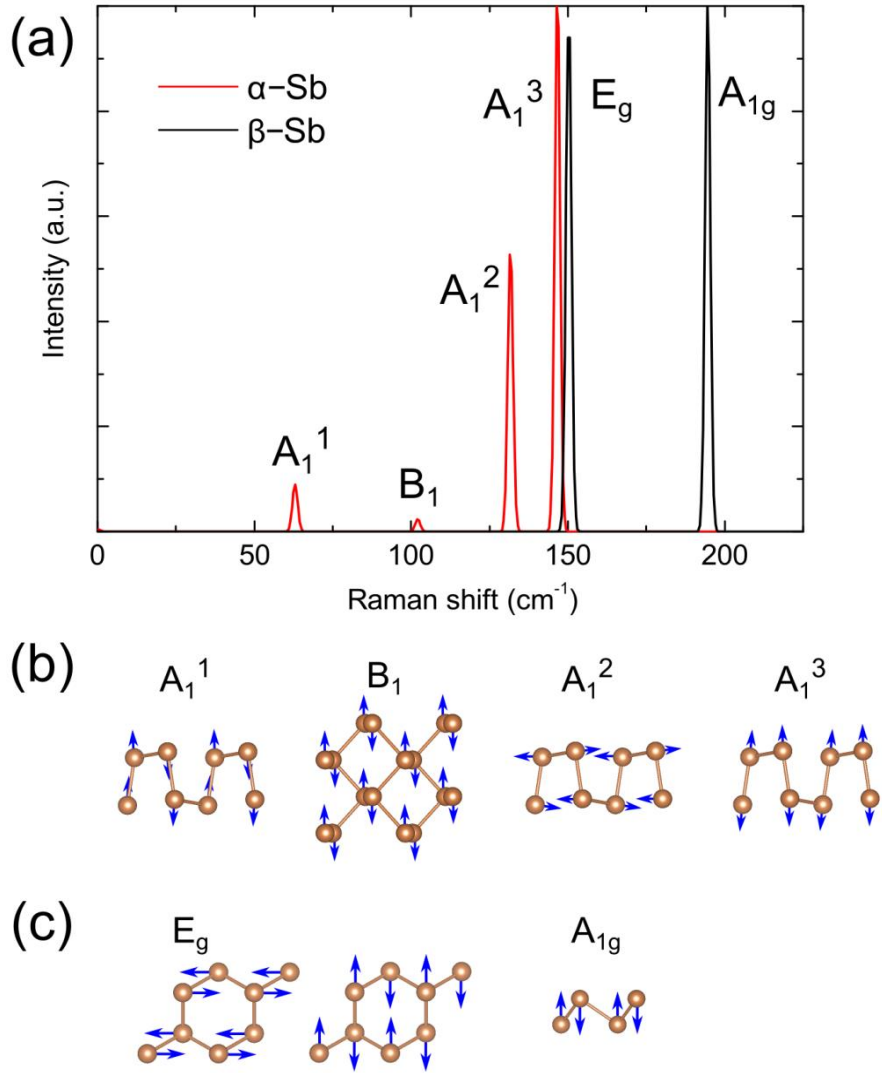

*Figure 3. The calculated Raman spectra (a) and the corresponding vibrational modes of antimonene allotropes (b and c).*

The calculated band structure, charge density and STM images are shown in

Figure 4. The *α*-Sb monolayer has a relatively small indirect band gap of ~0.28 eV. The valence band maximum (VBM) has a hybrid character of *s* orbitals and in-plane $p_x$ and $p_y$ orbitals (Figure S4, Supporting Information), which shows an almost linear dispersion at VBM. Due to the puckered structure, *α*-Sb has a stripe like STM surface characteristic (Figure 4(c)). The electronic band structure (Figure 4(d)) for the *β*-Sb monolayer shows it to be semiconducting with an indirect band gap of ~0.76 eV. A dot-like feature in the simulated STM image (Figure 4(f)) of the *β*-Sb monolayer results from its buckled surface.

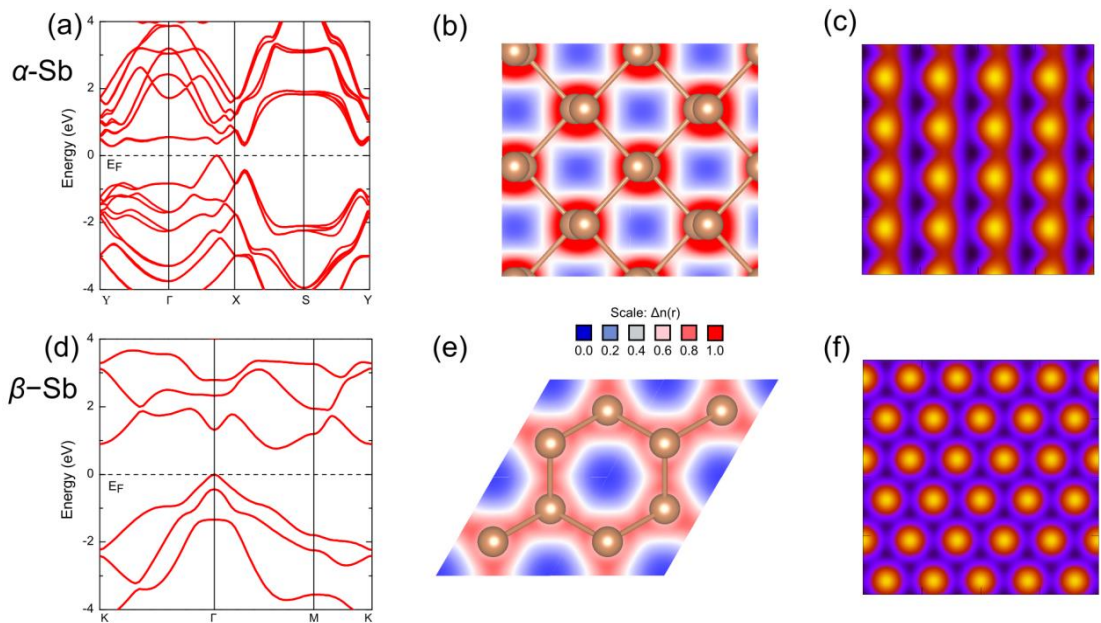

*Figure 4. Electronic properties of α-Sb (a, b, and c) and β-Sb (d, e, and f) monolayers: (a and d) band structure, (b and e) charge density projected in the plane, and (c and f) simulated STM images.*

Figure 5 shows atomic arrangements of antimonene multilayers. *β*-Sb multilayers prefers an ABC stacking similar to bulk Sb (Figure 5(d)); the AA-stacked bilayer is higher in energy by ≈24 meV/atom than the AB-stacked bilayer. The calculated interlayer distance is 3.65 Å. The band gaps of the bilayer and trilayer *β*-Sb (Figure S4 and S5, Supporting Information) decrease significantly due to the small surface states splitting as also predicted in the previous theoretical report on ultrathin *β*-Sb [21]. It is interesting to note that the binding energy of *β*-Sb bilayer is 124 meV/atom, which is much larger than that of other vdW layered materials, such as graphite (≈20 meV/atom [38]) and MoS$_2$ (≈60 meV/atom [38]). This is due to the partially overlapping of lone pair orbitals from the neighboring layers as seen from the charge density and the deformation change density plots given in Figures 5(e) and 5(f). Therefore, the mechanical exfoliation of bulk Sb is not expected to be relatively easier than that of graphite or MoS$_2$.

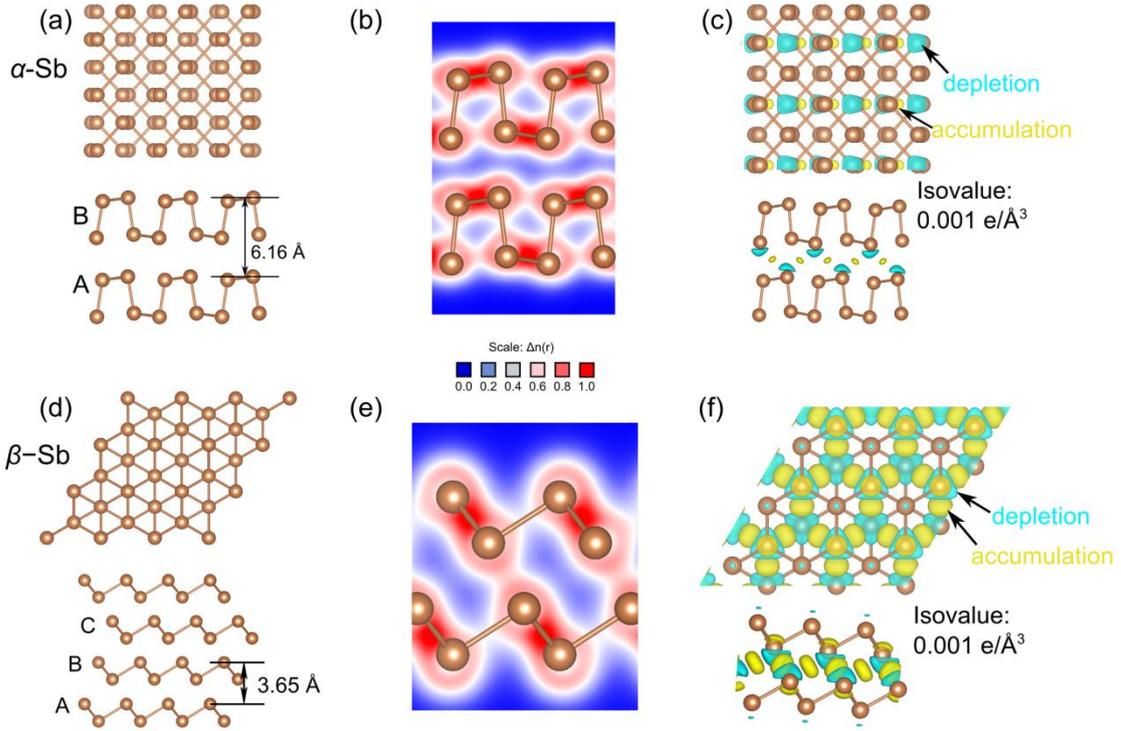

*Figure 5. α-Sb and β-Sb multilayers: (a) atomic structure, (b) charge density projected perpendicular to the layers, (c) deformation charge density for α-Sb bilayer; (d) atomic structure, (e) charge density projected perpendicular to the layers, and (f) the deformation charge density for β-Sb bilayer.*

*α*-Sb multilayers prefer AB stacking similar to that of black phosphorus (Figure 5(a)). The interlayer distance in *α*-Sb bilayer is calculated to be 6.16 Å. The binding energy of the AB-stacked *α*-Sb bilayer is calculated to be 68 meV/atom, which is close to those of other layered materials, e.g. MoS$_2$ [38]. The charge density in the interlayer region is very small (Figure 5(b)), and the electron redistribution in *α*-Sb bilayer (Figure 5(c)) is not as obvious as that in *β*-Sb bilayer. All these features indicate that the interlayer interaction is dominated by vdW dispersion in α-Sb multilayers. The AA-stacked bilayer is calculated to be 8 meV/atom higher in energy relative to the AB-stacked bilayer. The α-Sb bilayer and trilayer are calculated to be metallic with VBM and CBM crossing at the Fermi level (Figures S4 and S5, Supporting Information).

It is well known that strain can be introduced spontaneously by deposition of ultra-thin films on substrates with mismatched lattice constants. Application of strain to 2D atomic layers is also one of the possible approaches to tailor their electronic properties. Previous calculations on silicene, which has similar structure to β-Sb, have predicted it to sustain under the strain up to 20% [39-40]. Likewise, α-P shows superior mechanical properties due to its puckered structure, sustaining under the strain up to 30% along the armchair direction [41]. Experimentally, a large strain up to 30% could be applied to 2D materials by the use of stretchable substrates [42-43].

The tensile strain is defined as $\varepsilon = (a-a_0)/a_0$, where $a_0$ and $a$ are the lattice constants of the relaxed and strained structure, respectively. The stress-strain curve is calculated following the procedure of Wei and Peng [41] are shown in Figure 6. The stress is rescaled by the factor $Z/d$ to get the equivalent stress, where $Z$ is the cell length along $z$ direction, and $d$ is the interlayer spacing which is 3.65 and 6.16 Å for β-Sb and α-Sb, respectively. It should be noted that the interlayer distance predicted for β-P and α-P are 4.20 and 5.30 Å, respectively [6].

For α-Sb, the ideal strengths, which are defined as the maximum stress in the stress-strain curve, are ~10 GPa and ~4 GPa along the zigzag and armchair directions (Figure 6(a)). The corresponding critical strains are 18% and 32%. For β-Sb, the ideal strengths are ~10 and ~11 GPa along zigzag and armchair directions, respectively (Figure 6(b)). The corresponding critical strains are 15% (zigzag direction) and 18% (armchair direction). Both the ideal strength and critical strain are quite similar along the zigzag and armchair directions. This clearly shows that β-Sb has nearly isotropic mechanical properties while α-Sb exhibits strongly anisotropic mechanical characteristics. The critical strain along the armchair direction is extremely large in α-Sb, which will lead to strain engineering of its electronic properties.

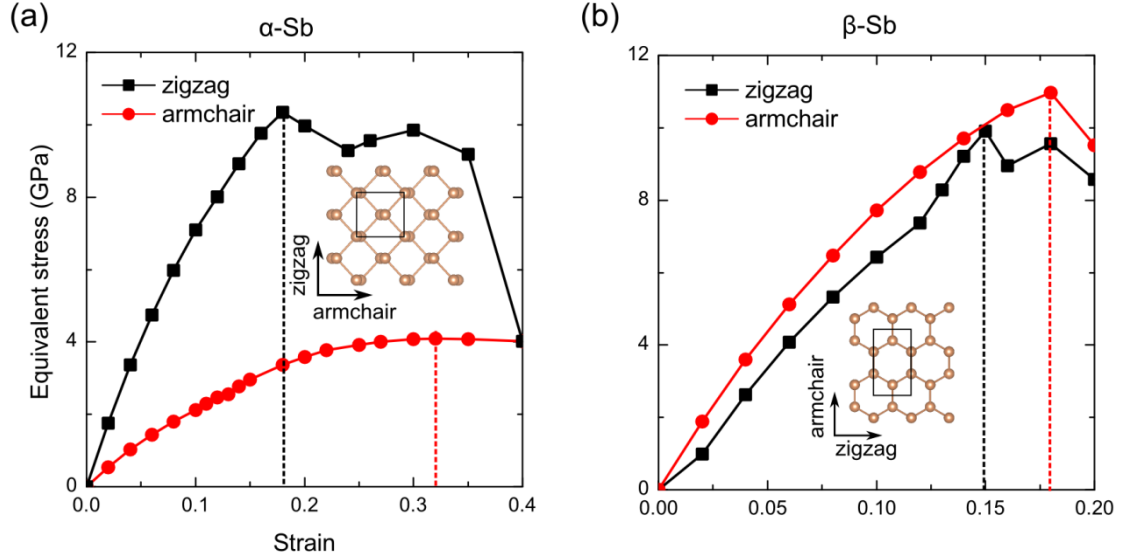

*Figure 6. Stress-strain relationship for antimonene monolayers: (a) α-Sb and (b) β-Sb.*

Next, we examine the tensile strain effect on the electronic properties of both Sb monolayers within the critical strain region. *α*-Sb has an indirect band gap and the tensile strain along the armchair direction induces an indirect-direct band gap transition (Figure 7(a)). With strain larger than 6%, a direct band gap at *V1* is predicted. For 11% strain, the band gap at *V1* decreases to 0.05 eV. Thereafter, the band gap gradually increases with strain larger than 11%, and reaches to 0.45 eV at 20% strain. For the tensile strain along zigzag direction (Figure 7(b)), CBM moves to *V2* point, and VBM moving to *Γ* for 8% strain. The strain induced indirect-direct band transition is mainly due to competition of states at *Γ*, *V1* and *V2*. Similar theoretical results have also been reported for *α*-P [44] and *α*-As layers [8].

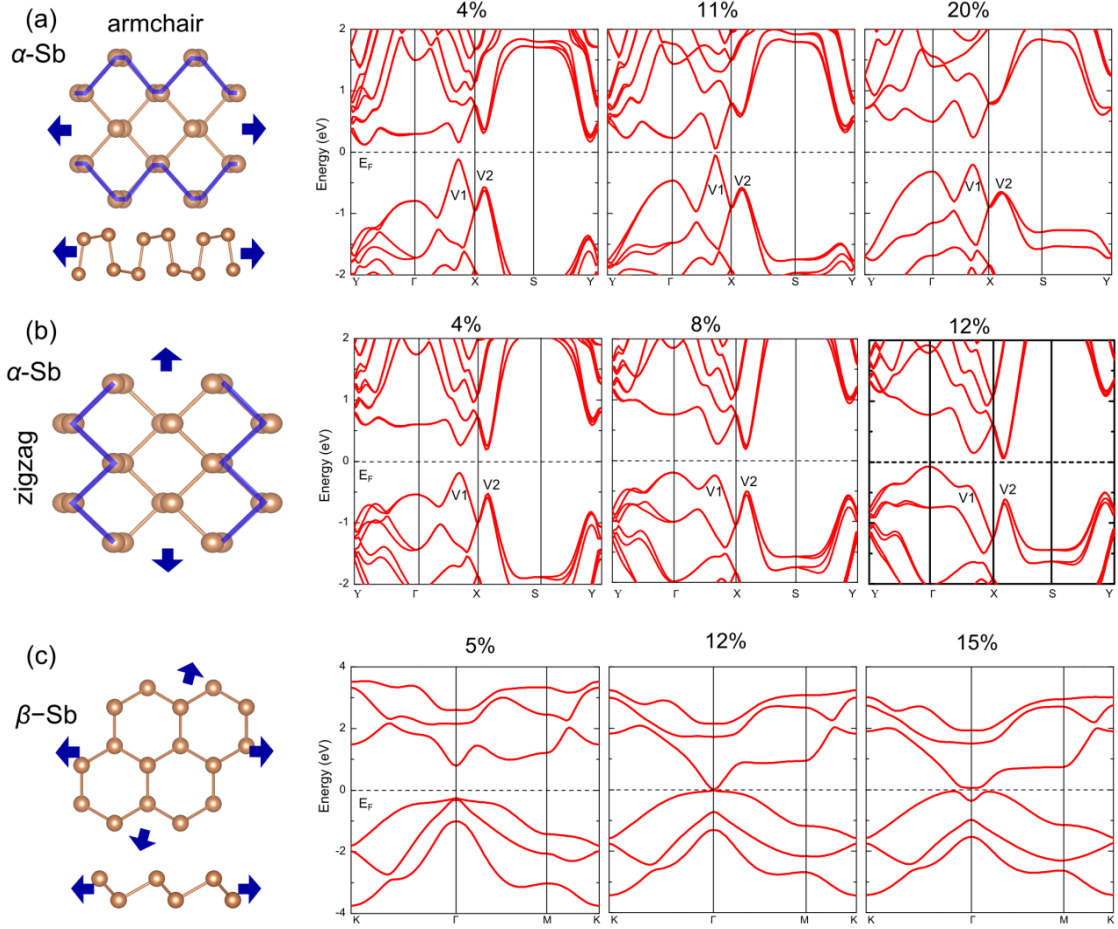

*Figure 7. Electronic band structures of α-Sb and β-Sb monolayers under various strains: (a) α-Sb under strain along armchair direction, (b) α-Sb under strain along zigzag direction, and (c) β-Sb under biaxial strain.*

Since *β*-Sb monolayer shows isotropic mechanical properties along the zigzag and armchair directions, a biaxial tensile strain was applied to the lattice as shown in Figure 7(c). *β*-Sb monolayer has (minimum) indirect band gap of 0.76 eV at the equilibrium configuration. Under 5% strain, its band gap becomes direct at *Γ*. The band gap closes under 12% strain, and reopens for strain 20%. Considering that *β*-Sb monolayer still preserves its buckled structure under 15% strain, its band gap can be effectively tuned by the in-plane strain.

Since the mechanical exfoliation (scotch tape) approach may not be the ideal approach to make antimonene, the standard chemical techniques are likely to play a major role in the synthesis of 2D antimonene system. For example, the ultrathin Bi(111) and Bi(110) films have been assembled on Si or highly ordered pyrolytic graphite (HOPG) substrate by vapor deposition [16, 45]. In the following work, we have considered an antimonene/graphene system taking graphene as substrate. The $α$-Sb/graphene system is simulated with a rectangular supercell of (4.5 Å×17.1 Å), and $β$-Sb/graphene is simulated with a parallelogram supercell of (10.6 Å×10.6 Å) as shown in Figure S6 (see Supporting Information). The choice of relatively large supercell essentially minimizes the lattice mismatch between graphene and antimonene to be less than 5%.

In the equilibrium configurations, the interlayer distance is larger than 3.2 Å at the LDA-DFT level of theory. The intra-planar (Sb-Sb) bond lengths are 2.82 and 2.86 Å in $α$-Sb/graphene, and 2.83 Å in $β$-Sb/graphene, which are similar to the calculated bond lengths in the freestanding monolayer. The calculated binding energies of $α$- and $β$-Sb on graphene substrate are 16 meV/atom and 14 meV/atom, respectively. The binding energy is defined in terms of energy difference between the constituent monolayers and the antimonene/graphene system. From the projected band structures (Figure S6, Supporting Information), the nature of the band gap of $β$-Sb monolayer is maintained on the graphene substrate, and a small charge transfer from $α$-Sb to graphene is calculated as also suggested by shifting of the Fermi level (Figure S6, Supporting Information). The calculated results therefore suggest that graphene could possibly serve as a substrate for the epitaxial growth of antimonene allotropes.

**SUMMARY**

In summary, DFT calculations are performed on 2D antimonene atomic layers. Our results show that $α$- and $β$-Sb monolayers are stable. Both monolayers are semiconductors with indirect band gap. $β$-Sb has nearly isotropic mechanical properties whereas $α$-Sb exhibits strongly anisotropic mechanical characteristics. Moderate tensile strain would induce indirect to direct band gap transition in antimonene. The calculated Raman spectrum prominently shows in-plane and out of plane vibrating modes that can be used to characterize antimonene monolayers.

**ASSOCIATED CONTENT**

**Supporting Information Available:**

Figure S1: Helmholtz free energy of phonon as a function of temperature for $\alpha$-Sb and $\beta$-Sb.

Figure S2: Cohesive energy of $\alpha$-Sb and $\beta$-Sb multilayers calculated at the LDA-DFT level of theory.

Figure S3: Simulated Raman spectrum for bulk Sb.

Figure S4: Density of states of multilayer antimonene: (a) $\alpha$-Sb, and (b) $\beta$-Sb.

Figure S5: Band structure of multilayer antimonene: (a) $\alpha$-Sb, and (b) $\beta$-Sb.

Figure S6: Structural and electronic properties of antimonene deposited on graphene: (a) and (b) $\alpha$-Sb/graphene; (c) and (d) $\beta$-Sb/graphene.

Table S1: Summary of the stability of group V elementary monolayers.

Table S2: The ground state structural parameters of $\alpha$- and $\beta$-Sb layers.

This material is available free of charge via the Internet at http://pubs.acs.org.

**ACKNOWLEDGEMENTS**

RAMA and Superior, high performance computing clusters at Michigan Technological University, were used in obtaining results presented in this paper. Supports from Dr. S. Gowtham are gratefully acknowledged. This research was partially supported by the Army Research Office through grant number W911NF-14-2-0088.

**TOC graphic**

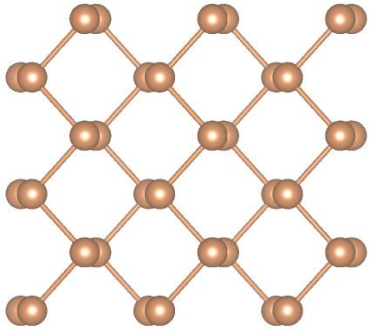 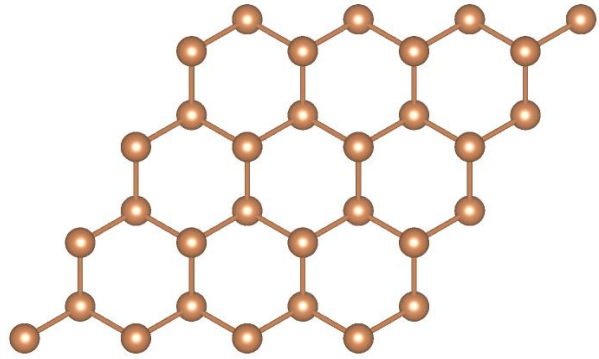

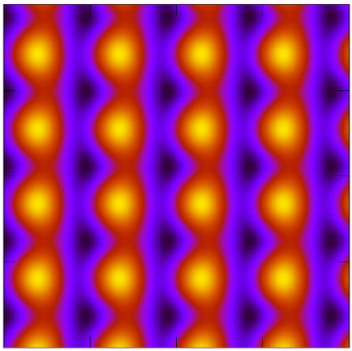 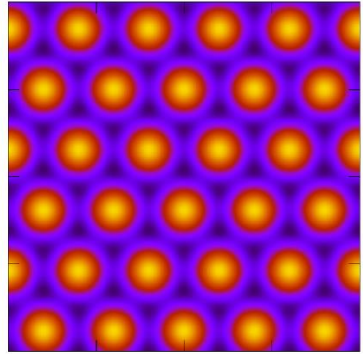

**Supporting Information**

**Atomically thin group-V elemental films: theoretical investigations of antimonene allotropes**

Gaoxue Wang[1], Ravindra Pandey[1]*, and Shashi P. Karna[2]

[1]Department of Physics, Michigan Technological University, Houghton, Michigan 49931, USA
[2]US Army Research Laboratory, Weapons and Materials Research Directorate, ATTN: RDRL-WM, Aberdeen Proving Ground, MD 21005-5069, USA

*Email: pandey@mtu.edu
   shashi.p.karna.civ@mail.mil

*Table S1. Summary of the stability of group V elementary monolayers in different phases. '√ (×)' means that the corresponding monolayer is stable (unstable), '-' means that the corresponding structure has not been investigated yet.*

|  | α | β | γ | δ |
|---|---|---|---|---|
| Phosphroene | √ (Ref [1-2]) | √ (Ref [3]) | √ (Ref [3]) | √ (Ref [3]) |
| Arsenene | √ (Ref [4-5]) | √ (Ref [5-6]) | - | - |
| Antimonene | √ (This work) | √ (Ref [7-8]) | × (This work) | × (This work) |
| Bismuthene | √ (Ref [9-11]) | √ (Ref [12-13]) | - | - |

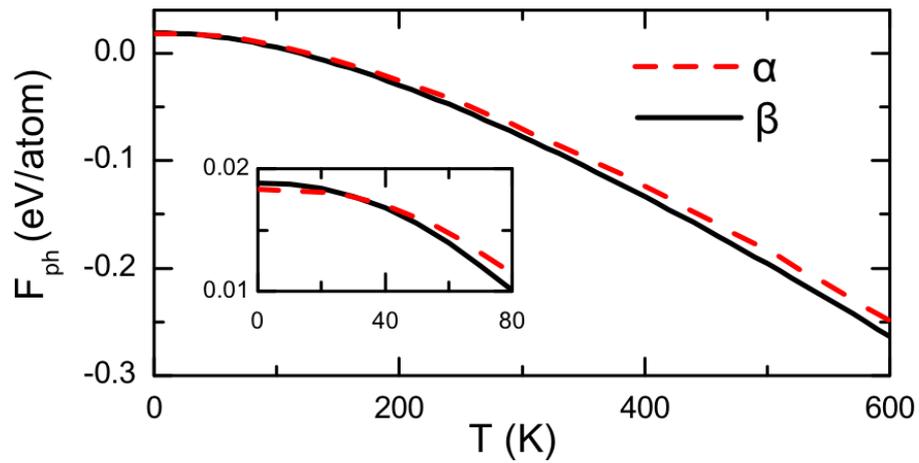

*Figure S1. Helmholtz free energy of phonon as a function of temperature for α-Sb and β-Sb.*

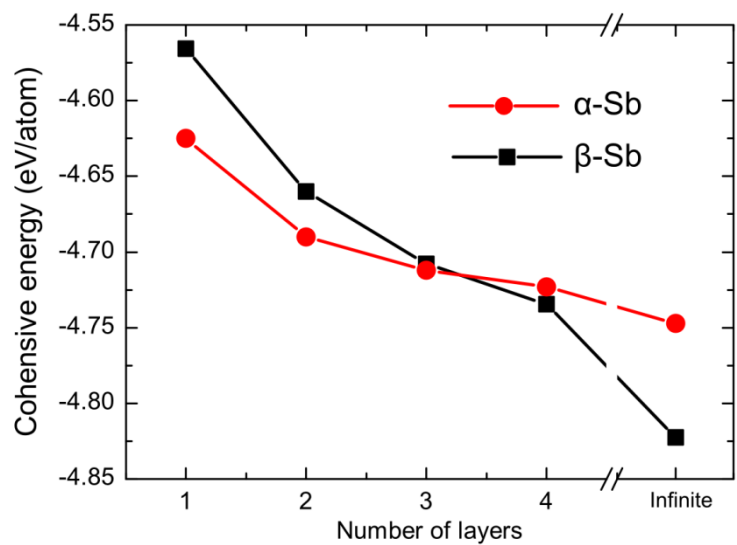

*Figure S2. Cohesive energy of α-Sb and β-Sb multilayers at the LDA-DFT level of theory.*

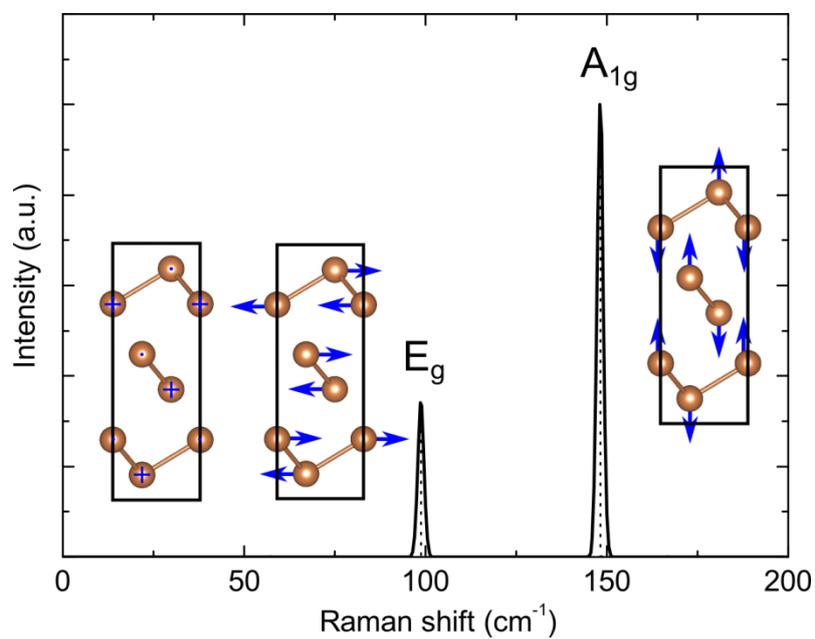

*Figure S3. Simulated Raman spectrum for bulk Sb. The insets show the vibrational modes at each peak.*

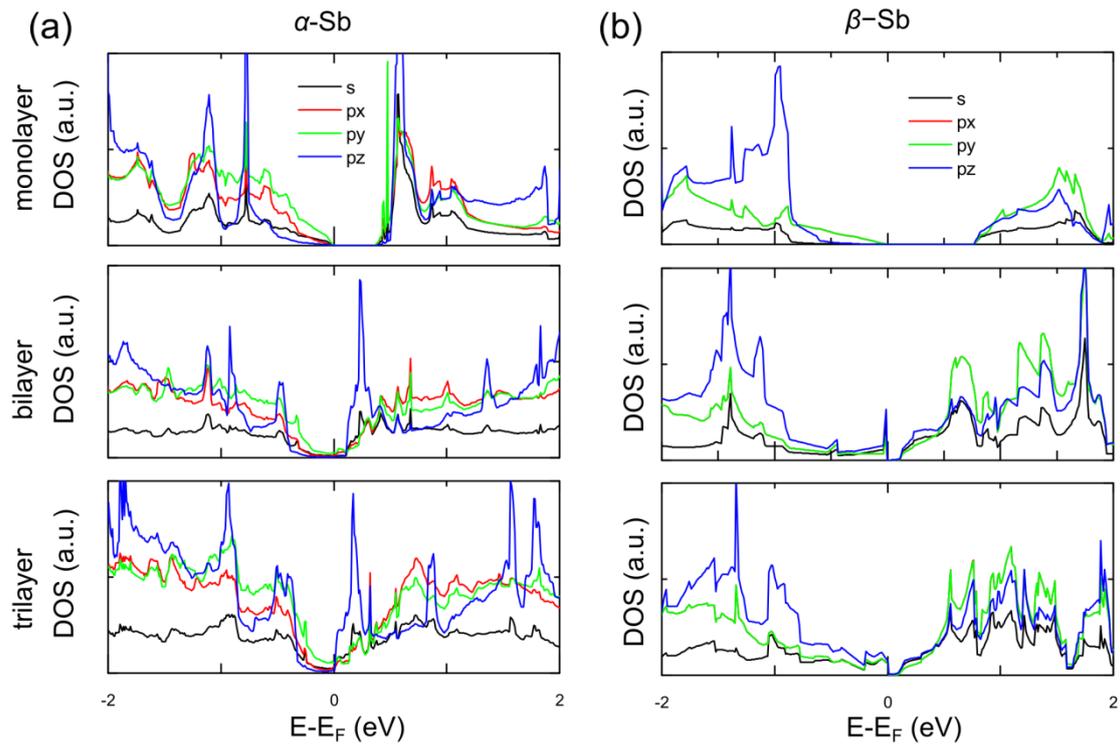

*Figure S4. Density of states of multilayer antimonene: (a) α-Sb, and (b) β-Sb.*

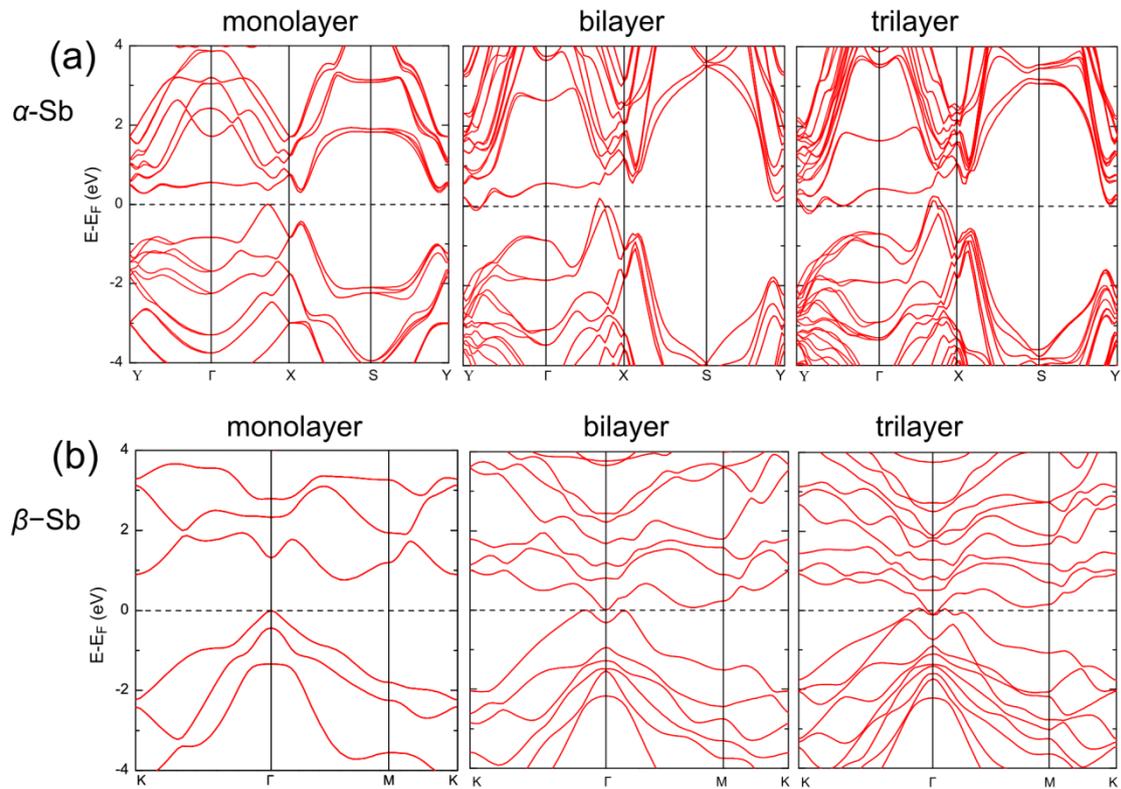

*Figure S5. Band structure of multilayer antimonene: (a) α-Sb, and (b) β-Sb.*

*Table S2. The ground state structural parameters of α- and β-Sb layers: a is the lattice constant, d is the interlayer distance, R is the near-neighbor distance at the LDA-DFT level of theory.*

|  | α-Sb | | | | | β-Sb | | |
| --- | --- | --- | --- | --- | --- | --- | --- | --- |
|  | $a_1$ (Å) | $a_2$ (Å) | $D$ (Å) | $R_1$ (Å) | $R_2$ (Å) | $a$ (Å) | $d$ (Å) | $R$ (Å) |
| Monolayer | 4.48 | 4.29 | - | 2.83 | 2.91 | 4.01 | - | 2.84 |
| Bilayer | 4.52 | 4.27 | 5.99 | 2.85 | 2.91 | 4.16 | 3.79 | 2.87 |
| Trilayer | 4.53 | 4.27 | 6.00 | 2.85 | 2.91 | 4.21 | 3.75 | 2.88 |
| Bulk | 4.66 | 4.31 | 6.16 | 2.88 | 2.91 | 4.31 | 3.65 | 2.91 |

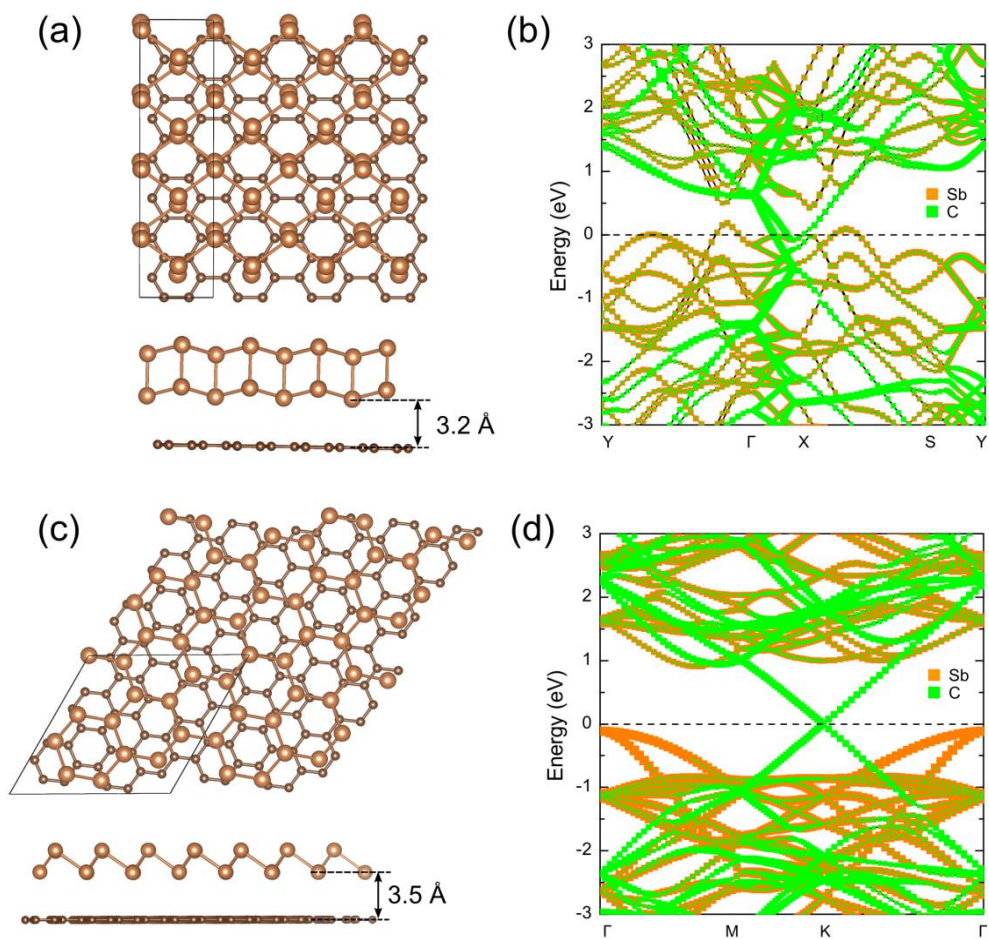

*Figure S6. The calculated structural and electronic properties of antimonene/graphene system: (a) and (b) α-Sb /graphene; (c) and (d) β-Sb /graphene.*